\documentclass[aps,prb,amsmath,amssymb,superscriptaddress,showpacs,twocolumn,floatfix]{revtex4}
\usepackage{amsfonts, relsize}
\usepackage{graphics}
\usepackage{graphicx}
\usepackage{hyperref}

\usepackage{color}
\begin{document}

\title{Migdal's theorem and electron-phonon vertex corrections in Dirac materials}
\author{Bitan Roy}
\affiliation{ Condensed Matter Theory Center, Department of Physics, University of Maryland, College Park, MD 20742, USA}

\author{Jay D. Sau}
\affiliation{ Condensed Matter Theory Center, Department of Physics, University of Maryland, College Park, MD 20742, USA}

\author{S. Das Sarma}
\affiliation{ Condensed Matter Theory Center, Department of Physics, University of Maryland, College Park, MD 20742, USA}

\date{\today}

\begin{abstract}
Migdal's theorem plays a central role in the physics of electron-phonon interactions in metals and semiconductors, and has been extensively studied theoretically for parabolic band electronic systems in three-, two-, and one-dimensional systems over the last fifty years.  In the current work, we theoretically study the relevance of Migdal's theorem in graphene and Weyl semimetals which are examples of 2D and 3D Dirac materials, respectively, with linear and chiral band dispersion. Our work also applies to 2D and 3D topological insulator systems. In Fermi liquids, the renormalization of the electron-phonon vertex scales as the ratio of sound ($v_s$) to Fermi ($v_F$) velocity, which is typically a small quantity. In two- and three-dimensional quasirelativistic systems, such as undoped graphene and Weyl semimetals, the one loop electron-phonon vertex renormalization, which also scales as $\eta=v_s/v_F$ as $\eta \rightarrow 0$, is, however, enhanced by an ultraviolet \emph{logarithmic divergent correction}, arising from the linear, chiral Dirac band dispersion. Such enhancement of the electron-phonon vertex can be significantly softened due to the logarithmic increment of the Fermi velocity, arising from the long range Coulomb interaction, and therefore, the electron-phonon vertex correction does not have a logarithmic divergence at low energy. Otherwise, the Coulomb interaction does not lead to any additional renormalization of the electron-phonon vertex. Therefore, electron-phonon vertex corrections in two- and three-dimensional Dirac fermionic systems scale as $v_s/v^0_F$, where $v^0_F$ is the bare Fermi velocity, and small when $v_s \ll v^0_F$. These results, although explicitly derived for the  intrinsic undoped systems, should hold even when the chemical potential is tuned away from the Dirac points. 
\end{abstract}

\pacs{63.22.-m, 81.05.ue, 04.50.-h}

\maketitle

\vspace{10pt}

\section{Introduction}

Two-dimensional honeycomb lattice, occupied by carbon atoms, offers an example of a novel state of matter, where the low energy excitations are described by two dimensional pseudorelativistic Dirac equation.\cite{wallace, semenoff} Conical dispersion of massless quasiparticles also emerges on the surface of strong $Z_2$ topological insulators,\cite{fu-kane} such as $Bi_2Se_3$, as well as topological crystalline insulators,\cite{liangfu_TCI} $SnTe$ for example.\cite{TCIexperiment} Despite such universal description at low energies, the microscopic origins of the quasirelativistic excitations in these materials are significantly different. While the lack of inversion symmetry in the honeycomb lattice produces the Dirac quasiparticles in graphene,\cite{HJR} the time-reversal symmetry and the mirror symmetry, respectively, give rise to such excitations on the surface of strong $Z_2$ and crystalline topological insulators.\cite{rmp-TI, liangfu_TCI, fu-kane} Recently, Weyl semimetals, described by three-dimensional massless Dirac equation, have also been realized in various noncentrosymmetric materials, such as $Cd_3As_2$,\cite{weylexperiment1} $Ni_3Bi$.\cite{weylexperiment2} In contrast to the conventional Fermi liquids, where low energy excitations live around a closed surface (Fermi surface), in all these systems the long-lived quasiparticles can only be found in the vicinity of few special points in the Brillouin zone, where the valence and the conduction bands touch each other. These are the so-called Dirac points, and the diverse materials containing these Dirac points are often classified together as Dirac materials.
\\

In addition to the fermionic degrees of freedom, smooth lattice deformations or motion of charge-neutralizing ionic background give rise to new excitations in the system, \emph{phonons}, which exist in all solid state systems as fundamental excitations of the lattice degrees of freedom. Phonon modes can also arise on the surface of topological insulators from its slowly varying deformations.\cite{thalmeier} Mutual couplings of these two degrees of freedom, the electron-phonon interactions, can lead to interesting many-body phenomena such as superconductivity.\cite{tinkham} Phonon driven pairing can also occur for chiral fermions, living on the surface of topological insulators, where the chemical potential naturally resides inside the conduction band, and in turn supports a finite Fermi surface.\cite{dassarma-li} Even though electron-phonon interactions can lead to significant renormalization of various single-particle quantities, such as electronic self-energy, in a seminal work, Migdal showed a long time ago that the correction or renormalization of the electron-phonon vertex due to the quantum fluctuations scales as $(m/M)^{1/2}$, where $m$ and $M$ are the electronic and ionic masses, respectively.\cite{migdal, schrieffer} Therefore, the renormalization of the electron-phonon vertex is typically small in Fermi liquids, since $m \ll M$. Although Migdal's original work dealt with three-dimensional Fermi liquids, later his work has been extended to two\cite{migdal2d} and one\cite{migdal1d} dimensional Fermi systems as well. On the other hand, the electron-phonon vertex correction can be significant, and consequently the Migdal's theorem breaks down, when the Debye frequency is comparable to the Fermi energy\cite{grimaldi-1}, or in the presence of van Hove singularities\cite{grimaldi-2}. Additionally, the higher order vertex corrections, which we neglect here, can also become significant at intermediate coupling\cite{bauer}. Hence, in the context of recently emerging two- and three-dimensional Dirac materials, a question arises quite naturally: Does the electron-phonon vertex renormalization in the quasirelativistic systems vanish or not as $\eta(=v_s/v_F) \rightarrow 0$? Or, in other words, does the Migdal's theorem for the smallness of the electron-phonon vertex renormalization hold for Dirac materials. This is the question that is addressed and answered in the current paper.
\\

In this paper, we wish to provide an answer to this question by taking into account the electron-phonon interactions in undoped (intrinsic) 2D graphene\cite{hwang-dassarma-phonon} and 3D Weyl semimetals. Our results are also applicable to the surface states of undoped topological insulators, when the chemical potential ($\mu$) is pinned at the Dirac point. We will also comment on the doped situation away from the Dirac point, showing that our results apply equally well to the doped situation.                      
\\

The rest of this paper is organized as follows. In the next section, we briefly discuss the electron-phonon vertex corrections and Migdal's theorem in three, two and one dimensional Fermi liquids with parabolic band dispersion in order to provide a context and notations for our new results on the Dirac materials. In Sec. III, we discuss the renormalization of the electron-phonon vertex in graphene. Section IV is devoted to the discussion of electron-phonon interaction and vertex corrections in Weyl semimetals. Our results are summarized in Sec. V. We relegate some details of the calculation to the Appendices. 
\\


\section{Migdal's theorem for non-relativistic fermions with parabolic band dispersion}
\begin{figure}[htb]
\includegraphics[width=5.5cm,height=4.5cm]{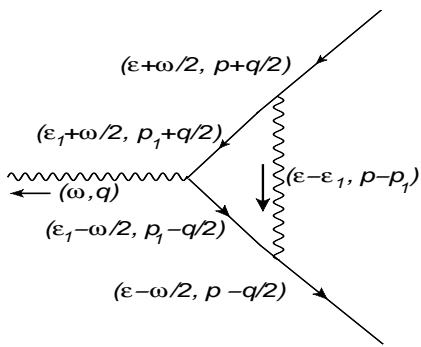}
\caption{One loop correction of the electron-phonon vertex. The wavy (solid) lines represent phonons (fermions).}
\end{figure}

We begin our discussion by reviewing the electron-phonon vertex renormalization for nonrelativistic fermions. Our derivation closely follows the original work of Migdal.\cite{migdal} The Hamiltonian describing the electron-phonon interaction has the form 
\begin{equation}
H=\sum_{k} \epsilon_k c^\dagger_k c_k + \sum_{q} \Omega_q\left( a^\dagger_{q} a_{q}\right)+H_{e-ph},
\end{equation}
where $c^\dagger_k$ and $c_k$ are, respectively, the fermionic creation and annihilation operator. The same quantities for phonons are represented by $a^\dagger_q$ and $a_q$, respectively. Phonon dispersion is given by $\Omega_q=v_s q$, where $v_s$ is the velocity of sound in the system. For simplicity the spin index of fermions is suppressed here. The electron-phonon interaction is represented by 
\begin{equation}
H_{e-ph}=\sum_{k,k',q} \alpha_q c^\dagger_k c_{k'} \left( a^\dagger_q+a_q\right) \delta_{k',k-q},
\end{equation}
where 
\begin{equation}\label{alphaFL}
\alpha_q = \lambda q \sqrt{\frac{\hbar}{\Omega_q \rho_m}},
\end{equation}
and $\rho_m$ is the ionic mass density in $d$ dimensions. Hereafter we set $\hbar=1$, and $\lambda \sim E_F$ (Fermi energy) is the deformation potential, coupling defining the basic electron-phonon interaction strength. The fermion and the phonon single-particle propagator is defined, respectively, by the Green's function $G_0$ and $D_0$ given below\cite{mahan}: 
\begin{equation}
G_0(\epsilon,p)=\frac{1}{\epsilon_p - \epsilon - i \Delta(p)},
D_0 (\omega,q)=\frac{2 \; \Omega_q}{\omega^2-\Omega^2_q + i \delta},
\end{equation} 
where $\delta, \Delta$ are the standard sign functions for fermionic and bosonic propagators. 
\\

The leading order correction to the electron-phonon vertex arises from the diagram shown in Fig.~1, and its contribution reads as 
\begin{eqnarray}
\Gamma_0(p,q) &=&\int_{\epsilon_1,\vec{p}_1} \alpha^2_{p-p_1}D_0(\epsilon-\epsilon_1, p-p_1) \; \nonumber \\
&\times&  G_0(\epsilon_1+\frac{\omega}{2},p_1+\frac{q}{2}) \: G_0(\epsilon_1-\frac{\omega}{2}, p_1-\frac{q}{2}).
\end{eqnarray}
After completing the frequency integral (over $\epsilon_1$), we obtain
\begin{eqnarray}
\Gamma_0=\frac{\lambda^2}{ \rho_m v_s}  \int_{\vec{p}_1} \frac{|\vec{p}-\vec{p}_1|}{\epsilon_{\vec{p}_1 + \frac{\vec{q}}{2}} -(\Omega_{p_1-p} + \frac{\omega}{2})-i \Delta(\Omega_{p_1-p} + \frac{\omega}{2})} \times
\nonumber \\
\hspace{-1cm}\frac{1}{\epsilon_{\vec{p}_1 - \frac{\vec{q}}{2}} -(\Omega_{p_1-p} - \frac{\omega}{2})-i \Delta(\Omega_{p_1-p} - \frac{\omega}{2})} + {\cal O}(E^{-2}_F).
\end{eqnarray}
In performing the integral over $\epsilon_1$, we have ignored the terms coming from the poles of the fermion Green function $G_0$, since these are small in the limit $\Omega_q \ll E_F$. Next we linearize the spectrum around the Fermi surface. Then, $\int p^{d-1} dp \to \int \rho(E) dE$, where $\rho(E)=p^{d-1}/v_F$ is the density of states, and $v_F$ (typically $\gg v_s$) is the Fermi velocity. For the parabolic dispersion in two dimensions, the density of states $\rho(E)=k_F/v_F$ is independent of $E$, where $k_F$ is Fermi momentum. Assuming $\rho(E)$ to be a constant near the Fermi energy, one can write the vertex correction as
\begin{eqnarray}\label{gamma0afterfreq}
\Gamma_0 = \hat{\lambda}_{FL,d} (v_s q) \: \int d\Omega_\theta \frac{1}{v_F q \cos{\theta}-\omega + i \delta sign(\omega)} \nonumber \\
\int^{\infty}_{-\infty} dE \bigg[\frac{1}{E + \frac{v_F q \cos{\theta}}{2} -\frac{\omega}{2}-i \Delta(\Omega_{k_F} -E_F + \frac{\omega}{2})}  \nonumber \\
  - \frac{1}{E - \frac{v_F q \cos{\theta}}{2}-\frac{\omega}{2} -i \Delta(\Omega_{k_F} -E_F - \frac{\omega}{2})} \bigg],
\end{eqnarray}
where $\hat{\lambda}_{FL,d}=\frac{\lambda^2 k^{d-1}_F}{v_F \rho_m v^2_s}$ is the dimensionless electron-phonon coupling in $d$-dimensional Fermi liquid. The integral over $E$ leads to a combination of $\Theta$ functions, $\big[ \Theta\big(\Omega_{k_F}-E_F+\frac{\omega}{2} \big)$ $-\Theta\big(\Omega_{k_F}-E_F-\frac{\omega}{2} \big) \big]$, which can yield only a factor of unity.
\\

Upon completing the angular integrations, and assuming $\omega \sim v_s q \ll v_F q$, the electron-phonon vertex correction in $d=3, 2$ becomes 
\begin{eqnarray}\label{vercorrectionFL}
\Gamma_0 = \hat{\lambda}_{FL,d} \bigg[ \frac{v_s}{v_F} \left(\frac{v_s}{v_F} + i \pi \right) \bigg] \equiv \hat{\lambda}_{FL,d} \; \eta \left(\eta+i \pi \right),
\end{eqnarray}
where $\eta=v_s/v_F$. In one dimension there is no angular integral, and for $v_F \gg v_s$, the electron-phonon vertex correction is simply $\Gamma_0= \eta \hat{\lambda}_{FL,1}$. Therefore, when the velocity of sound ($v_s$) is much smaller than the Fermi velocity ($v_F$), which is typically the situation in solid state materials, the correction to the electron-phonon vertex is always suppressed by a factor $v_s/v_F$, and can therefore be neglected in two and three dimensions. In one dimension, however, this result is true only if we neglect the backscattering, which leads to Peierls instability at weak interactions.\cite{giamarchi} We do not discuss the rather special (and pathological) 1D case any more in this work, concentrating entirely on 2D and 3D systems with our focus being on graphene (2D) and Weyl semimetals (3D).
\\

Notice, we here present the electron-phonon vertex correction as a function of the ratio $v_s/v_F$, which is slightly different from the results quoted in standard text books, stating $\Gamma_0\sim (m/M)^{1/2}$.\cite{schrieffer, fetter-walecka} These two results are, however, completely equivalent. On the other hand, upon casting $\Gamma_0$ as a function of $v_s/v_F$, we can compare the electron-phonon vertex corrections in Fermi liquids with the ones in graphene and Weyl semimetals, which we discuss next. In fact, expressing the electron-phonon vertex function  in terms of a ratio of the characteristic phonon (sound) velocity and the electron (Fermi) velocity is the appropriate theoretical (as well as physical) approach, since this result is more general than introducing an unnecessary ionic (or electronic) mass into the formalism.    
\\


\section{Electron-phonon interaction and Migdal's theorem in graphene}

We now consider the interaction of low energy Dirac quasiparticles in half-filled (intrinsic) undoped graphene with in-plane 2D acoustic phonon. Let us define a two component spinor $\Psi^\top= \big[ u(\vec{K}+\vec{p}),$ $v(\vec{K}+\vec{p}) \big]$, where $u$ and $v$ are the fermionic annihilation operators on two triangular sublattices of the honeycomb lattice, and $\vec{K}$ corresponds to the Dirac point.\cite{wallace, semenoff} In this basis the noninteracting Hamiltonian with only nearest-neighbor hopping in graphene takes the form $\hat{H}_D(p)=v_F \left( \sigma_1 p_x +\sigma_2 p_y \right)$, at low energies ($|\vec{p}| \ll |\vec{K}|$), where $\sigma_{1,2}$ are the standard two-dimensional Pauli matrices. For simplicity, we here suppress the valley and spin degrees of freedom of Dirac fermions since acoustic phonons, which we are considering in the current paper, do not couple the valleys or spins. The coupling of Dirac fermions with acoustic phonons is given by $H_{e-ph}=\alpha_q \left(\Psi^\dagger \Psi\right) \Phi$, where $\Phi$ represents phonon field, and $\alpha_q$ is defined in Eq. (\ref{alphaFL}).\cite{hwang-dassarma-phonon, Khveshchenko} Hereafter, we set $\hbar=1$.
\\

The Euclidean action corresponding to the electron-phonon interaction in $d-$dimensions reads as $S=\int d^d\vec{x} d\tau {\cal L}$, where $\tau$ is the imaginary time and 
\begin{eqnarray}\label{action}
{\cal L} =  \Psi^\dagger \big[ \partial_\tau + \hat{H}_D \big] \Psi + \Phi^\dagger 
\big[ \partial_\tau + v_s q \big] \Phi + H_{e-ph}. 
\end{eqnarray}
Scale invariance of the effective Euclidean action manifests the scaling dimensions of various quantities as follow: $[x]=-1, [\tau]=-z$, yielding $[\Psi]=[\Psi^\dagger]=d/2=[\Phi]=[\Phi^\dagger]$, where $z$ is the dynamical critical exponent, which in our problem is \emph{unity}. The Fermi and the sound velocity scales as $[v_F]=[v_s]=z-1$. The electron-phonon vertex $\alpha_q \sim \sqrt{q}$, and its scaling dimension is $[\alpha_q]=z-d/2-1/2$. Therefore, the bare electron-phonon coupling is \emph{irrelevant} in graphene ($d=2$), and the one loop correction to the electron-phonon vertex scales $\sim \Lambda$, where $\Lambda$ is the ultraviolet cutoff for the conical dispersion. 
\\

Next we evaluate the one loop renormalization of the electron-phonon vertex in graphene (see Fig.~1), which reads as
\begin{eqnarray}\label{gamma0}
\Gamma_0&=& \int \frac{d\nu}{2 \pi} \frac{d^2\vec{p}}{(2 \pi)^2} \alpha^2_{p} D_0(\nu,p) G_0(\omega-\nu,k-p) \nonumber \\
&\times& G_0(\omega+\alpha-\nu, k+q-p),
\end{eqnarray}   
where 
\begin{equation}
D_0(\nu,p)=\frac{2 v_s p}{\nu^2+ (v_s p)^2}, G_0 (\nu,p)= \frac{-i \nu+\hat{H}_D(p)}{\nu^2+ v^2_F p^2}.
\end{equation}
Notice that the momentum routing here is slightly different than that in the previous section (consult first paragraph of Appendix A). We here only present the central results, but details of the calculation can be found in Appendix A. Combining the denominators in Eq. (\ref{gamma0}) using the Feynman parameters $(x,y)$ and completing the frequency integral over $\nu$, we obtain
\begin{eqnarray}\label{gamma01afterfrequency}
\Gamma_0 = \frac{ \lambda^2}{8 \rho_m}\int^1_0 dx dy \int \frac{d^2 \vec{p}}{(2 \pi)^2} 
\bigg[ - \frac{p^2}{[\Delta]^{3/2}} + 3 \frac{p^2 F}{[\Delta]^{5/2}}\bigg],
\end{eqnarray}
where $\Delta(=\Delta_1+\Delta_2)$ and $F$ are lengthy functions of frequencies, momenta and two velocities ($v_F$, $v_s$), shown in Eqs.(\ref{delta1mudelta2}) and (\ref{Ffunction}), respectively. 
\\

\begin{figure}[htb]
\includegraphics[width=4.0cm,height=3.0cm]{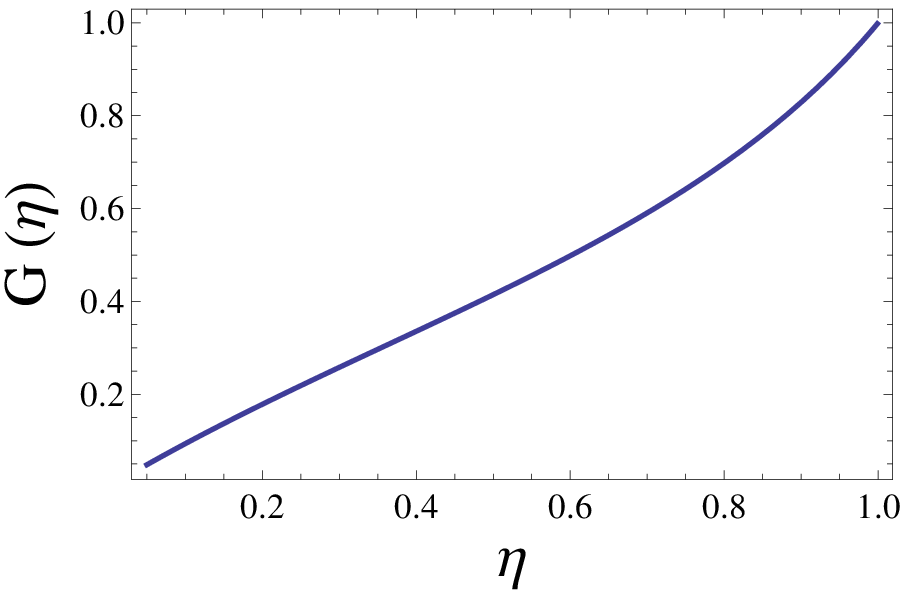}
\includegraphics[width=4.0cm,height=3.0cm]{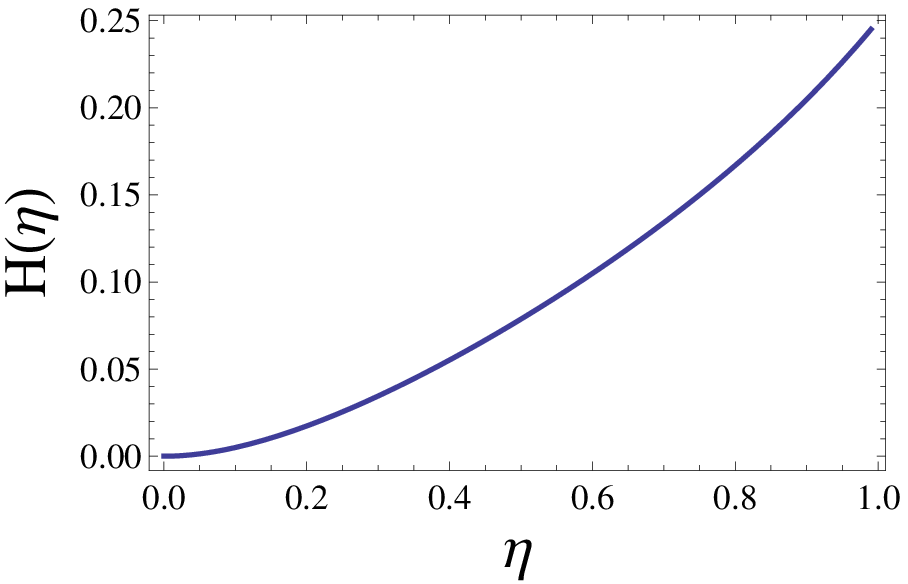}
\caption{(Color online) $G(\eta)$(left) and $H(\eta)$(right) as functions of $\eta=v_s/v_F$.}
\end{figure}

Next we evaluate the momentum integral over $p$, after promoting it from physical two dimensions to $d$ dimensions, where $d=1+\epsilon$, since the electron-phonon interaction ($\alpha_q$) is \emph{marginal} for $d=1$. This procedure allows us to capture the divergent terms in $\Gamma_0$ quite efficiently. Otherwise, with $\beta=(1-x-y) v^2_s+(x+y)v^2_F$, the total divergent contribution in $\Gamma_0$ goes as
\begin{eqnarray}\label{gama10divdimreg}
\Gamma^{div}_0 &=& \frac{\lambda^2}{16\pi \rho_m}  \int^1_0 dx dy \bigg[ -\frac{1}{\beta^{3/2}}+\frac{3 v^2_F}{\beta^{5/2}}\bigg] \Gamma(-\frac{\epsilon}{2}) \nonumber \\
&=& \bigg[ \frac{\lambda^2 \Lambda}{4 \pi v_F \rho_m v^2_s} \bigg] \log{\left(\frac{\Lambda}{\tilde{\Lambda}}\right)} G(\eta)
\equiv \hat{\lambda}_{D,2} \log{\left(\frac{\Lambda}{\tilde{\Lambda}}\right)} G(\eta), \nonumber \\
\end{eqnarray}
where $\hat{\lambda}_{D,2}=\frac{\lambda^2 \Lambda}{4 \pi v_F \rho_m v^2_s}$ is the dimensionless electron-phonon coupling in graphene, and for $\eta (=v_s/v_F)<1$
\begin{equation}\label{Gfunction}
G(\eta)= \frac{\eta\left(1-\eta (2-\eta^2)^{-1/2} \right)}{1-\eta^2}.
\end{equation}
The final expression in Eq.~(\ref{gama10divdimreg}) is obtained by rewriting the divergent term $\Gamma(-\frac{\epsilon}{2})=$ $\Lambda^\epsilon \log(\Lambda/\tilde{\Lambda})$ with $\epsilon=1$, in terms of the ultraviolet cutoff $\Lambda \sim 1/a$, where $a$ is the lattice spacing in graphene and $\tilde{\Lambda} (\ll \Lambda)$ $\sim \sqrt{k^2+q^2+k \cdot q}$ or $(\frac{|\omega|}{v_F}+\frac{|\alpha|}{v_s})$ (whichever is larger) in $d=2$. Variation of $G(\eta)$ with $\eta$ is shown in Fig.~2 (left), and as $\eta \rightarrow 0$, $G(\eta) \rightarrow \eta+{\cal O}(\eta^2)$. Comparing $\hat{\lambda}_{D,2}$ and $\hat{\lambda}_{FL,2}$, we find that the ultraviolet cutoff $\Lambda$ for the conical Dirac dispersion plays the role of Fermi momentum $k_F$ of the nonrelativistic fermions. The leading order finite term in $\Gamma_0$ as $k,q,\omega \rightarrow 0$, while the frequency of the incoming phonon ($\alpha$) is kept finite, reads as $\Gamma^{fin}_0=\hat{\lambda}_{D,2} H(\eta) \big( \frac{\alpha}{v_F \Lambda}\big)$. The velocity dependent function $H(\eta)$ is shown in Fig.~2 (right), and as $\eta \rightarrow 0$, $H(\eta) \rightarrow (2-\sqrt{2}) \eta^2 + {\cal O}(\eta^3)$. However, $\Gamma^{fin}_0$ vanishes as we send the ultraviolet cutoff $\Lambda \to \infty$, and does not contribute to the renormalization of the electron-phonon vertex, and $\Gamma_0 = \Gamma^{div}_0$.
\\

Therefore, the renormalization of the electron-phonon vertex $\alpha_q$ develops a {\em logarithmic divergent contribution} [see Eq.~(\ref{gama10divdimreg})], and as $\eta \rightarrow 0$ it becomes $\Gamma_0=\eta \hat{\lambda}_{D,2} \log{(\Lambda/\tilde{\Lambda})}$. The extra logarithmic divergence in the Dirac system can be understood in the context of the derivation of Migdal's theorem in the parabolic band electronic systems (presented in Sec. II) in terms of the difference in the density of states $\rho(E)$ between graphene and systems with a Fermi surface. While the density of states in two dimension systems with parabolic dispersion is independent of energy [as assumed in Eq.(\ref{gamma0afterfreq})], the density of states in the Dirac system vanishes linearly as the energy ($E$) goes to zero at the Dirac point. Introducing such a linear $E$ dependence in the density of states at the Fermi energy allows us to obtain the same logarithmic divergent result from Eq. (\ref{gamma0afterfreq}) as we have obtained for the conical dispersion. 
\\

 If we take the long range Coulomb interaction in graphene into account, the Fermi velocity ($v_F$) also increases logarithmically as 
\begin{equation}
v_F = v^0_F \left[ 1+ \alpha_{FS} \log\left(\frac{\Lambda}{\Lambda_e} \right) \right], 
\end{equation}
where $\alpha_{FS}$ is the fine-structure constant, $\Lambda_e$ is the electron momentum, and $v^0_F$ is the bare Fermi velocity.\cite{son-FV, HWDSTS-FV} Otherwise, in two spatial dimensions electronic charge does not get renormalized due to the nonanalytic nature of the Coulomb propagator $(\sim 1/|q|)$. Consequently, the velocity dependent function $G(\eta)\rightarrow \eta+{\cal O} (\eta^2)$ is suppressed at least by a factor $\log(\Lambda/\Lambda_e)$. Hence, the logarithmic enhancement of Fermi velocity softens the logarithmic divergence in $\Gamma_0$, and the electron-phonon vertex correction in graphene scales as $\eta_0=v_s/v^0_F$, which is precisely consistent with  Migdal's theorem, and is small when $\eta_0$ is small, as in ordinary parabolic band metallic Fermi liquids.  We note, however, that if electron-electron interaction is neglected and only electron-phonon interaction is taken into account, then the vertex correction to the electron-phonon coupling at the Dirac point develops an ultraviolet divergence as the renormalization group flow goes to lower energy. This is an inevitable feature of all Dirac materials, implying that although Migdal's theorem remains formally valid, the coupling develops an ultraviolet logarithmic divergence at the Dirac point.  
\\

Coulomb interaction, in principle, can lead to additional renormalization of the electron phonon vertex \cite{aleiner-basko}, which reads as 
\begin{eqnarray}\label{coulomb}
\Gamma^{C}_0&=& e^2 \int \frac{d\nu}{2 \pi} \frac{d^2\vec{p}}{(2 \pi)^2} D^{C}_0(\nu,p) G_0(\omega-\nu,k-p) \nonumber \\
&\times& G_0(\omega+\alpha-\nu, k+q-p),
\end{eqnarray}
where $D^{C}_0(\nu,p)=1/|p|$ is the Coulomb propagator, and $e$ is electronic charge. The only term that remains finite as we send the ultraviolet cutoff $\Lambda \to \infty$ is obtained simply by setting $k,q,\omega,\alpha=0$ in the above expression. However, this term  
\begin{equation}\label{coulombfreq}
\Gamma^{C}_0= e^2 \int \frac{d^2\vec{p}}{(2 \pi)^2} D^C_0 (\nu,p) \int^{\infty}_{-\infty} \frac{d\nu}{2 \pi} \frac{-\nu^2+ v^2_F p^2}{(\nu^2+v^2_F p^2)^2} \equiv 0,
\end{equation}
dictating that Coulomb interaction by itself does not renormalize the electron-phonon vertex in graphene. But, Coulomb interaction among the electrons themselves produces an ultraviolet divergence in the bare Fermi velocity which precisely cancels the ultraviolet enhancement arising in the electron-phonon vertex, maintaining the strict validity of Migdal's theorem in graphene.


\section{Migdal's theorem in Weyl semimetal}

Next we evaluate the renormalization of the electron-phonon vertex in 3D Weyl semimetals. The massless Dirac Hamiltonian in three dimensions takes the form $\hat{H}_D= v_F\big(\sigma_1 p_x$ $+ \sigma_2 p_y +\sigma_3 p_z\big)$, and the electron-phonon coupling is $H_{e-ph}=\alpha_q \left(\Psi^\dagger \Psi \right) \Phi$. Scaling dimension of $\alpha_{q}$ (see previous section) dictates that the bare electron-phonon vertex is {\em irrelevant} in Weyl semimetals ($d=3$), and one loop correction of $\alpha_q$ scales as $\Lambda^2$. 
\\

The vertex correction is once again given by Eq.~(\ref{gamma0}), with a three-dimensional momentum integral. However, by virtue of computing the momentum integral in general $d$-spatial dimensions, we can immediately extract the leading ultraviolet divergent term in $\Gamma_0$ in terms of $\Lambda$ and $\tilde{\Lambda}$ in Weyl semimetals from Eq.~(\ref{gama10divdimreg}), since $\Gamma(-\frac{\epsilon}{2})=\Lambda^2 \log(\Lambda/\tilde{\Lambda})$ in $d=3$, yielding
\begin{eqnarray}
\Gamma_0= \frac{\Lambda^2 \lambda^2}{4 \pi v_F \rho_m v^2_s} G(\eta) \log{\left(\frac{\Lambda}{\tilde{\Lambda}}\right)}
\equiv \hat{\lambda}_{D,3} G(\eta) \log{\left(\frac{\Lambda}{\tilde{\Lambda}}\right)}, \nonumber \\
\end{eqnarray}
where $\hat{\lambda}_{D,3}=\frac{\Lambda^2 \lambda^2}{4 \pi v_F \rho_m v^2_s}$ is the dimensionless electron-phonon coupling in Weyl semimetals. We here neglect the contribution from the finite terms, since they all vanish upon setting the ultraviolet cutoff $\Lambda \to \infty$. 
\\

Hence, the electron-phonon vertex correction for three-dimensional Weyl semimetals, which as $\eta(=v_s/v_F) \to 0$ reads as $\Gamma_0= \eta \hat{\lambda}_{D,3} \log(\Lambda/\tilde{\Lambda})$, also develops an ultraviolet logarithmic divergent contribution. However, in three dimensions, due to the analytic structure of the Coulomb propagator $(\sim 1/q^2)$, both the electronic charge and the Fermi velocity get renormalized, which can be captured from the following renormalization group flow equations \cite{goswami-chakravarty, hosur-parameswar-ashwin, nagaosa, rosenstein}
\begin{equation}
\frac{d \alpha_{FS}}{d \log{(\Lambda/\Lambda_e)}}=-\frac{4 \alpha^2_{FS}}{3 \pi}, \: \:
\frac{d v_F}{d \log{(\Lambda/\Lambda_e)}}= v_F \frac{2 \alpha_{FS}}{3 \pi}.
\end{equation}
Solution of these two coupled flow equations is shown in Fig.~3. Therefore, the long range Coulomb interaction in Weyl semimetals also produces a logarithmic enhancement of the Fermi velocity, similar to graphene. Thus the logarithmic divergence in $\Gamma_0$ gets suppressed even in $d=3$. As a result, the electron-phonon vertex correction scales as $\eta_0=v_s/v^0_F$, where $v^0_F$ is bare Fermi velocity in Weyl semimetals. Thus, Migdal's theorem remains formally valid in the 3D Weyl semimetals when $v_s \ll v^0_F$, just as in 2D graphene.    
\\

\begin{figure}[htb]
\includegraphics[width=8.0cm,height=5.0cm]{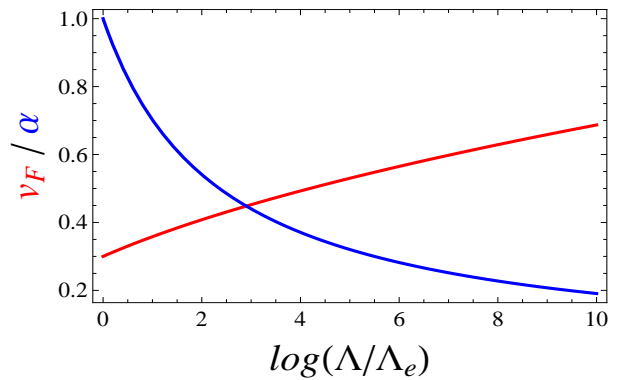}
\caption{(Color online) Solutions of the renormalized Fermi velocity $v_F$ (red) and fine structure constant $\alpha$ (blue), as a function of momentum cutoff $\log{(\Lambda/\Lambda_e)}$, for a particular choice of initial values $v^0_F=0.3$ and $\alpha^{0}_{FS}=1$. }
\end{figure}

Renormalization of the electron-phonon vertex in Weyl semimetals, arising from the Coulomb interaction, is given by Eq. (\ref{coulomb}), with $D^{C}_0(\nu,p)=1/|p|^2$. However, the term that survives when we send the cutoff $\Lambda \to \infty$ is \emph{zero}, as shown in Eq. (\ref{coulombfreq}). Therefore, the electron-phonon vertex in Weyl semimetals does not get renormalized due to the Coulomb interaction, similar to the corresponding situation in graphene. 
\\


\section{Summary and Discussion}

To summarize, we here address the relevance of electron-phonon vertex correction in graphene and Weyl semimetals. In two- and three- dimensional Fermi liquids the vertex correction at one loop level scales as the ratio of sound ($v_s$) to Fermi velocity ($v_F$), which is typically a small quantity. In one-dimensional Fermi gas, a similar conclusion holds only if we neglect the effect of backscattering, which leads to Peierls instability at arbitrarily weak couplings.\cite{giamarchi} Taking into account the coupling of linearly dispersing acoustic phonons\cite{hwang-dassarma-phonon, Khveshchenko} with massless Dirac fermions in graphene ($d=2$) and Weyl semimetals ($d=3$), we show that renormalization of electron-phonon vertex is $\Gamma_0 =\eta \hat{\lambda}_{D,d} \log(\Lambda/\tilde{\Lambda})$, as $\eta(=v_s/v_F) \rightarrow 0$, where $\hat{\lambda}_{D,d}\left(=\frac{\lambda^2 \Lambda^{d-1}}{4 \pi v_F \rho_m v^2_s} \right)$ is the dimensionless electron-phonon coupling in the $d$-dimensional Dirac system. Here, $\Lambda$ corresponds to the ultraviolet cutoff for the conical dispersion of Dirac quasiparticles, and $\tilde{\Lambda}$ is a combination of electron and phonon momenta/frequencies. Therefore, the electron-phonon vertex corrections in graphene and Weyl semimetals suffer ultraviolet {\em logarithmic divergent correction}, which, however, gets suppressed due to the {\em logarithmic enhancement} of the Fermi velocity, arising from the long range Coulomb interaction between the electrons themselves, yielding $v_F \sim v^0_F \log(\Lambda/\Lambda_e)$, where $v^0_F$ is the {\em bare} Fermi velocity and $\Lambda_e \sim$ electron momentum.\cite{son-FV, HWDSTS-FV, goswami-chakravarty, hosur-parameswar-ashwin, nagaosa, rosenstein} The Coulomb interaction otherwise does not renormalize the electron-phonon vertex in graphene or Weyl semimetals. Hence, the vertex correction in two- and three- dimensional Dirac fermionic systems scales as $v_s/v^0_F$, and Migdal's theorem, associated with the smallness of the electron-phonon vertex corrections in Fermi liquids, is also valid for pseudorelativistic systems, at least when $v_s \ll v^0_F$.
\\

It is interesting to contrast electron-electron and electron-phonon interaction effects on graphene with respect to the ultraviolet logarithmic divergence arising from the large momentum cutoff inherent in the chiral, linear band dispersion of Dirac materials.  It is well known that the ultraviolet divergence does not affect the electron-electron interaction vertex correction, affecting only the self-energy function, and thus leading to a logarithmic velocity renormalization\cite{HWDSTS-FV} and consequently to a logarithmic running of the graphene effective fine structure constant characterizing the electron-electron interaction strength.  By contrast, electron-phonon interaction does not affect the self energy of graphene\cite{dassarma-li} while affecting the electron-phonon vertex function through the logarithmic ultraviolet factor. Including both effects to the leading order, however, the divergent logarithm drops out of the vertex correction restoring the usual Migdal's theorem, since the ultraviolet logarithmic term cancels out, as we argue above.  We believe that this result of ours, although derived in the leading order, remains valid to all orders in the electron-phonon vertex function as can be verified by a dimensional power counting.
\\

We expect that our results hold even for doped (extrinsic) graphene and Weyl semimetals, as well as for the surface states of topological insulators, where the chemical potential ($\mu$) is inside the conduction band. At finite doping electron-phonon interactions give rise to both intraband as well as interband excitations. The ultraviolet divergent term in the vertex correction $\Gamma_0$, arising from interband excitations, is similar to the $\mu=0$ situation, and scales as $ (v_s/v_F)\log(\Lambda/k_F)$, where $k_F =\mu/v_F$. Such divergent correction can again be compensated by the similar logarithmic enhancement of the Fermi velocity due to the electron-electron interactions.\cite{DSHW-FV} Otherwise, the intraband excitations in the vicinity of the Fermi pocket give rise to a finite contributions, as in regular Fermi liquids. Hence, at finite chemical potential, the one-loop electron-phonon vertex correction is also expected to scale as $v_s/v^0_F$, and remain small as long as $v_s \ll v^0_F$. Thus, Migdal's theorem remains formally valid in Dirac materials, both in the doped and the undoped situation.       
\\

Finally, we mention that although we have worked with a particular model of electron-phonon interaction, namely, the deformation potential coupling (which is always present in any solid state materials involving a lattice, including 2D\cite{hwang-dassarma-phonon} and 3D\cite{dassarma-li} Dirac materials) in this paper, other possible models of electron-phonon coupling would give the same formal result about the formal validity of Migdal's theorem in Dirac materials. We provide in Appendix B our results for an alternative model of electron-phonon coupling in graphene,\cite{felix} which is not very physical, but nevertheless obeys Migdal's theorem.
\\


\acknowledgements

This work is supported by US-ONR and by LPS-CMTC.

\appendix

\section{Detail of electron-phonon vertex correction for Dirac fermions}

We here provide some details of the calculation of electron-phonon vertex correction ($\Gamma_0$) in graphene and Weyl semimetals, presented in Secs. III and IV, respectively. The expression of $\Gamma_0$ in Eq.~(\ref{gamma0}) arises from Fig.~1, by using a slightly different momentum and frequency routing, where we take $(\omega,q)\rightarrow(\alpha,q)$, $(\epsilon-\epsilon_1,p-p_1) \rightarrow (\nu,p)$, $(\epsilon \pm \omega/2, p \pm q/2) \rightarrow (\omega_T+\omega_{\pm}, k_T+k_\pm)$, $(\epsilon_1 \pm \omega/2, p_1 \pm q/2) \rightarrow (\omega_T+\omega_{\pm}-\nu, k_T+k_\pm-p)$, with $\omega_T=(\omega+\alpha)/2$, $k_T=(k+q)/2$, $\omega_\pm=(\omega \pm \alpha)/2$, $k_\pm=(k\pm q)/2$. In this notation $\nu$ and $p$ are the internal frequency and momentum (hence these are the integral variables) respectively, as shown in Eq.~(\ref{gamma0}).
\\

To evaluate $\Gamma_0$, first we need to combine the denominators of Eq.~(\ref{gamma0}) and cast it as a symmetric function of internal frequency $\nu$. This can be achieved by introducing two Feynman parameters ($x,y$) as follows\cite{peskin-schroeder}:
\begin{widetext}
\begin{equation}
\frac{1}{\nu^2+ v^2_s p^2} \times \frac{1}{(\omega-\nu)^2+ v^2_F (k-p)^2} \times \frac{1}{(\omega+\alpha-\nu)^2+ v^2_F (k+q-p)^2} \nonumber
\end{equation}
\begin{equation}
=2 \int^1_{0} dx \; dy \;
\bigg[(1-x-y)(\nu^2+ v^2_s p^2) + x ((\omega-\nu)^2+ v^2_F (k-p)^2)+ y ((\omega+\alpha-\nu)^2+ v^2_F (k+q-p)^2) \bigg]^{-3}. 
\end{equation}
\end{widetext}
Taking $\nu-(x+y)\omega - y \alpha \rightarrow \nu$, the above expression reads as 
\begin{equation}
2 \int^1_0 dx \; dy \; \big[ \nu^2 + \Delta_1 + \Delta_2 \big]^{-3},
\end{equation}
where 
\begin{eqnarray}
\Delta_1 = [(1-x-y) v^2_s + (x+y) v^2_F]p^2 - 2 v^2_F(x+y) (k \cdot p) \nonumber \\
- 2 v^2_F y (q \cdot p) + (x+y) v^2_F k^2+ y^2 v^2_F q^2 + 2 y v^2_F (k \cdot q), \nonumber
\end{eqnarray}
\begin{eqnarray}\label{delta1mudelta2}
\Delta_2 &=& (x+y) (1-x-y) \omega^2 + y(1-y) \alpha^2  \nonumber \\
&+& 2 y (1-x-y) \omega \alpha.
\end{eqnarray}
 After performing the same change of variable in the numerators of Eq. (\ref{gamma0}), a set of standard integrals 
\begin{eqnarray}
\int^\infty_{-\infty} \frac{d \nu}{2 \pi} \: \frac{\nu^2}{[\nu^2+ \Delta]^3}&=& 
\frac{1}{16} \: \frac{1}{[\Delta]^{3/2}},\nonumber \\
\int^\infty_{-\infty} \frac{d \nu}{2 \pi} \: \frac{1}{[\nu^2+ \Delta]^3}&=&
\frac{3}{16} \: \frac{1}{[\Delta]^{5/2}},
\end{eqnarray}  
where $\Delta=\Delta_1+\Delta_2$, leads to Eq.(\ref{gamma01afterfrequency}), with 
\begin{eqnarray}\label{Ffunction}
 F = \big[-i [(x+y-1)\omega+ y \alpha] + v_F \sigma_j {(k-p)}_j \big] \times \nonumber \\ 
 \big[-i [(x+y-1)\omega+ (y-1) \alpha] + v_F \sigma_j {(k+q-p)}_j \big]. 
\end{eqnarray}
Summation over the repeated index ($j$) is assumed throughout. 
\\

Next we need to perform the integration over the internal momentum $p$. In order to evaluate the integrals over $p$, we first express the function $\Delta$ in the denominators of Eq.(\ref{gamma01afterfrequency}) as a rotationally symmetric function of $p$. It can be accomplished by rewritting 
\begin{eqnarray}\label{Deltamanipulation}
[\Delta]^{n} &=& [\beta]^{n} \bigg[ \big( p-a k -b q \big)^2+a (1-a) k^2 \nonumber \\
&+& b (1-b) q^2+ 2 b(1-a)(k \cdot q)+ \frac{\Delta_2}{\beta}\bigg]^n,
\end{eqnarray}  
where $a=\frac{v^2_F(x+y)}{\beta}$, $b=\frac{v^2_F y}{\beta}$, and $\beta=(1-x-y)v^2_s+(x+y) v^2_F$. Upon taking $p-a k -b q \rightarrow p$, $\Delta$ becomes rotationally symmetric in $p^2$. Performing the same shift of variable in the numerators of Eq.~(\ref{gamma01afterfrequency}), we can separate it into two categories: $(i)$ terms that give rise to ultra-violet divergences, which for two entries in Eq. (\ref{gamma01afterfrequency}) 
\begin{eqnarray}
I_1=\int \frac{d^2 \vec{p}}{(2 \pi)^2} \frac{p^2}{[\Delta]^{3/2}}, \quad
I_2=\int \frac{d^2 \vec{p}}{(2 \pi)^2} \frac{p^2 F}{[\Delta]^{5/2}},
\end{eqnarray}
are given by
\begin{eqnarray}
I^{div}_1 &=& \frac{1}{\beta^{3/2}} \int \frac{d^dp}{(2 \pi)^d} \frac{p^2}{(p^2+\tilde{\Delta})^{3/2}}
=\frac{1}{\beta^{3/2}} \frac{1}{2 \pi} \Gamma(-\frac{\epsilon}{2}), \nonumber \\
I^{div}_2 &=& \frac{v^2_F}{d \beta^{5/2}} \int \frac{d^dp}{(2 \pi)^d} \frac{p^4}{(p^2+\tilde{\Delta})^{5/2}}
=\frac{v^2_F}{\beta^{5/2}} \frac{1}{2 \pi} \Gamma(-\frac{\epsilon}{2}), \nonumber \\
\end{eqnarray}   
respectively, yielding Eq. (\ref{gama10divdimreg}), where 
\begin{eqnarray}
\tilde{\Delta}=a(1-a) k^2 + b (1-b) q^2+ 2 b(1-a)(k \cdot q)+ \frac{\Delta_2}{\beta}. \nonumber
\end{eqnarray}
$(ii)$ The rest of the terms give finite contributions, which, however, vanish as we send the ultraviolet cutoff $\Lambda \rightarrow \infty$. Otherwise, the leading order finite term in $\Gamma_0$, as $k,q,\omega \rightarrow 0$ is given by $\Gamma^{fin}_0$, shown in Sec. III of the paper.   
\\

\section{Vertex correction of alternative electron-phonon interactions in graphene}

In graphene there are other models for electron-phonon interactions,\cite{felix} given by 
\begin{equation}\label{ephnew}
\tilde{H}_{e-ph}=g q \sqrt{\frac{\hbar}{\rho_m \Omega_q}} \sum_{j=1,2} \hat{d}_{j}\left(\Psi^\dagger \sigma_j \Psi\right) \Phi,
\end{equation} 
where $(\hat{d}_1,\hat{d}_2)=\left( \hat{q}^2_x-\hat{q}^2_y, 2 \hat{q}_x \hat{q}_y\right)$. Such electron-phonon interactions arise from the modification of the nearest-neighbor hopping amplitudes in honeycomb lattice. In an alternative approach, the effect of the modulated hopping can be captured by introducing time-reversal symmetric axial gauge potentials, coupled minimally to the Dirac fermions in graphene.\cite{roy-herbut-pseudo} For simplicity, we here only consider the longitudinal phonon mode. Renormalization of the above electron-phonon coupling ($g_q$) is 
\begin{eqnarray}
 \Gamma_l = \sigma_l \int \frac{d\nu}{2 \pi} \frac{d^2p}{(2 \pi)^2} D_0(\nu,p) \bigg[ \alpha^2_p G_0(\omega-\nu,k-p) \nonumber \\
 \sigma_l G_0(\omega+\alpha-\nu,k+q-p) + g^2_p \sum_{j=1,2} \hat{d}_{j} \sigma_j \nonumber \\
 G_0(\omega-\nu,k-p) \sigma_l G_0(\omega+\alpha-\nu,k+q-p) \hat{d}_j \sigma_j \bigg],
\end{eqnarray}  
where $l=1$ or $2$, and $g_q=g q/\sqrt{\rho_m \Omega_q}$ (after setting $\hbar=1$). Due to the underlying rotational symmetry of the system, one finds $\Gamma_1=\Gamma_2=\bar{\Gamma}$(say). Following the steps shown in Sec.~III and Appendix A, upon completing the frequency integral we obtain 
\begin{eqnarray}
\bar{\Gamma}&=&\frac{\lambda^2}{8 \rho_m} \int^{1}_0 dx dy \int \frac{d^2p}{(2 \pi)^2}\bigg[ - \frac{p^2}{[\Delta]^{3/2}} + 3\frac{p^2 \tilde{F}}{[\Delta]^{5/2}}\bigg] \nonumber \\
&+& \frac{g^2}{8 \rho_m} \int^{1}_0 dx dy \int \frac{d^2p}{(2 \pi)^2} \left(\hat{d}_{1} \sigma_1 -\hat{d}_{2} \sigma_2 \right) \nonumber \\
&\times& \bigg[ - \frac{p^2}{[\Delta]^{3/2}} + 3\frac{p^2 \tilde{F}}{[\Delta]^{5/2}}\bigg] \left(\hat{d}_{1} \sigma_1 +\hat{d}_{2} \sigma_2 \right), 
\end{eqnarray} 
where
\begin{eqnarray}
\tilde{F}= \big[-i [(x+y-1)\omega+ y \alpha] - v_F (-1)^{j} \sigma_j (k-p)_j \big] \nonumber \\
\big[-i [(x+y-1)\omega+ (y-1) \alpha] + v_F \sigma_j (k+q-p)_j \big].
\end{eqnarray}
Using the integral identities 
\begin{equation}
\int \frac{d^2p}{(2 \pi)^2} \frac{p^2_1-p^2_2}{(p^2+\tilde{\Delta})} \equiv 0, 
\int \frac{d^2p}{(2 \pi)^2} \frac{p_1p_2}{(p^2+\tilde{\Delta})} \equiv 0,
\end{equation}
it can be immediately shown that the divergent contributions from all the terms in the last equations, except the first term, is precisely zero. Therefore, renormalization of the electron-phonon vertex $g_q$ is given by
\begin{eqnarray}
\bar{\Gamma} &=& \frac{\lambda^2 H(\eta)}{4 \pi v_F \rho_m v^2_s } \Gamma(-\frac{\epsilon}{2}) \equiv \hat{\lambda}_{D,2} H(\eta) \log\left(\frac{\Lambda}{\tilde{\Lambda}} \right) \nonumber \\
&\rightarrow& \hat{\lambda}_{D,2} (2-\sqrt{2}) \eta^2 \log\left(\frac{\Lambda}{\tilde{\Lambda}}\right),
\end{eqnarray}
as $\eta(=v_s/v_F) \to 0$. Hence, the one loop renormalization of the electron-phonon vertex $g_q$ also suffers ultraviolet divergent logarithmic correction. However, the logarithmic enhancement of the Fermi velocity ($v_F$), arising from the electron-electron (Coulomb) interaction, suppresses such enhancement of electron-phonon vertex. Consequently, the renormalization of the electron-phonon vertex $g_q$ scales as $v^2_s/(v^0_F)^2$, and Migdal's theorem remains valid when $v_s \ll v^0_F$. 
\\

Coulomb interaction also leads to renormalization of the electron-phonon vertex $g_q$, given by 
\begin{eqnarray}
\bar{\Gamma}^C &=& e^2 \sigma_l \int \frac{d\nu}{2\pi} \frac{d^2 p}{(2 \pi)^2} D^C_0(\nu,p) G_0(\omega-\nu,k-p) \sigma_l \nonumber \\
&\times& G_0(\omega+\alpha-\nu,k+q-p),
\end{eqnarray}
with $l=1$ or $2$. Upon setting the ultraviolet cutoff ($\Lambda$) to infinity, the vertex correction due to the Coulomb interaction goes as
\begin{eqnarray}
\bar{\Gamma}^C &=&-e^2 \int \frac{d^2p}{(2\pi)^2} \int^\infty_{-\infty} \frac{d\nu}{2\pi} \frac{-\nu^2+v^2_F(p^2_1-p^2_2)}{(\nu^2+p^2)^2} \nonumber \\
&=& \left( \frac{e^2}{8 \pi v_F} \right) \: \log{\left( \frac{\Lambda}{\Lambda_e}\right)}.
\end{eqnarray}
Therefore, correction to the electron-phonon vertex, arising from the Coulomb interaction has ultraviolet logarithmic correction. Nevertheless, logarithmic enhancement of the Fermi velocity once again suppresses such logarithmic divergent correction of the vertex renormalization, and $\bar{\Gamma}^C \sim \frac{e^2}{8 \pi v^0_F}$, where $v^0_F$ is the bare Fermi velocity. Therefore, the electron-phonon vertex correction $\bar{\Gamma}^C$ is small as long as the bare dimensionless Coulomb coupling or the bare fine structure constant in graphene $e^2/(8 \pi v^0_F) \ll 1$.          
\\

The electron-phonon interactions $g_q$ also provide additional renormalization ($\tilde{\Gamma}_0$) of the original electron-phonon coupling $\alpha_q$ in graphene, introduced in Sec. III, where
\begin{eqnarray}
&& \tilde{\Gamma}_0 = \sum_{j=1,2}\int \frac{d\nu}{2 \pi} \frac{d^2p}{(2 \pi)^2} g^2_q D_0(\nu,p) \hat{d}_{j} \sigma_j G_0(\omega-\nu,k-p) \nonumber \\
&&  G_0(\omega+\alpha-\nu, k+q-p) \hat{d}_{j} \sigma_j.
\end{eqnarray}  
After some straightforward algebra, as shown in Sec. III and Appendix A, we obtain 
\begin{eqnarray}
\tilde{\Gamma}_0= \hat{g} G(\eta) \log \left(\frac{\Lambda}{\tilde{\Lambda}} \right) 
\rightarrow \hat{g} \eta \log \left(\frac{\Lambda}{\tilde{\Lambda}} \right),
\end{eqnarray}
as $\eta(=v_s/v_F) \to 0$, where $\hat{g}=\frac{g^2 \Lambda}{4 \pi v_F \rho_m v^2_s}$ is another dimensionless electron-phonon coupling in graphene. Taking the enhancement of the Fermi velocity due to the Coulomb interaction into account, we find that $\tilde{\Gamma}_0$ scales as $v_s/v^0_F$, and Migdal's theorem remains valid when $v_s \ll v^0_F$.


\begin{thebibliography}{99}
\bibitem{wallace}  P. R. Wallace, Phys. Rev. {\bf 71}, 622 (1947).
\bibitem{semenoff} G. W Semenoff, Phys. Rev. Lett. {\bf 53}, 2449 (1984).
\bibitem{fu-kane} L. Fu, C. L. Kane, Phys. Rev. B {\bf 76}, 045302 (2007).
\bibitem{liangfu_TCI} L. Fu, Phys. Rev. Lett. {\bf 106}, 106802 (2011).
\bibitem{TCIexperiment} Y. Tanaka, Z. Ren, T. Sato, K. Nakayama, S. Souma, T. Takahashi, K. Segawa, and Y. Ando, Nat. Phys. {\bf 8}, 800 (2012).
\bibitem{HJR} I. F. Herbut, V. Juri\v{c}i\'{c}, and B. Roy, Phys. Rev. B {\bf 79}, 085116 (2009). 
\bibitem{rmp-TI} M. Z. Hassan, and C. L. Kane, Rev. Mod. Phys. {\bf 82}, 3045 (2010); X. L. Qi, and S. C. Zhang, {\em ibid}, {\bf 83}, 1057 (2011).  
\bibitem{weylexperiment1} S. Borisenko, Q. Gibson, D. Evtushinsky, V. Zabolotnyy, B. Buechner, and R. J. Cava, arXiv:1309.7978.
\bibitem{weylexperiment2} Z. K. Liu, B. Zhou, Z. J. Wang, H. M. Weng, D. Prabhakaran, S.-K. Mo, Y. Zhang, Z. X. Shen, Z. Fang, X. Dai, Z. Hussain, and Y. L. Chen, Science, {\bf 343}, 864 (2014).
\bibitem{thalmeier} P. Thalmeier, Phys. Rev. B {\bf 83}, 125314 (2011).
\bibitem{tinkham} M. Tinkham, \emph{Introduction to superconductivity}, Dover, New York, 1996. 
\bibitem{dassarma-li} S. Das Sarma, and Q. Li, Phys. Rev. B {\bf 88}, 081404(R) (2013). 
\bibitem{migdal} A. B. Migdal, Sov. Phys. JETP, {\bf 7}, 996 (1958).
\bibitem{schrieffer} See also J. R. Schrieffer, {\em Theory of Superconductivity} (Benjamin-Cummings, New York, 1983).
\bibitem{migdal2d} A. Madhukar, Solid State Commun. {\bf 24}, 11 (1977).
\bibitem{migdal1d} M. Apostol, and I. Baldea, J. Phys. C: Solid State Phys. {\bf 15}, 3319 (1982).
\bibitem{grimaldi-1} L. Pietronero, S. Str\"{a}ssler, and C. Grimaldi, Phys. Rev. B {\bf 52}, 10516 (1995);  C. Grimaldi, L. Pietronero, and S. Str\"{a}ssler, Phys. Rev. Lett. {\bf 75}, 1158 (1995).
\bibitem{grimaldi-2} E. Cappelluti and L. Pietronero, Phys. Rev. B {\bf 53}, 932 (1996).
\bibitem{bauer} J. Bauer, J. E. Han, and O. Gunnarsson, Phys. Rev. B {\bf 84}, 184531 (2011). 
\bibitem{hwang-dassarma-phonon} E. H. Hwang, and S. Das Sarma, Phys. Rev. B {\bf 77}, 115449 (2008).
\bibitem{mahan} G. D. Mahan, {\em Many-particle Physics}, 3rd ed. (Kluwer Academic/Plenum, New York, 2000).
\bibitem{giamarchi} T. Giamarchy, {\em Quantum Physics in One Dimension} (Oxford University Press, New York, 2004). 
\bibitem{fetter-walecka} A. L. Fetter, and J. D. Walecka,{\em Quantum Theory of Many-Particle Systems} (Dover, New York, 2003).
\bibitem{Khveshchenko} D. V. Khveshchenko, J. Phys.: Cond. Matt. {\bf 21}, 075303 (2009).
\bibitem{son-FV} D. T. Son, Phys. Rev. B {\bf 75}, 235423 (2007).
\bibitem{HWDSTS-FV} S. Das Sarma, E. H. Hwang, and W-K. Tse, Phys. Rev. B {\bf 75}, 121406 (2007).
\bibitem{aleiner-basko} D. M. Basko, I. L. Aleiner, Phys. Rev. B {\bf 77}, 041409(R) (2008).
\bibitem{goswami-chakravarty} P. Goswami, and S. Chakravarty, Phys. Rev. Lett. {\bf 107}, 196803 (2011).
\bibitem{hosur-parameswar-ashwin} P. Hosur, S. A. Parameswaran, and A. Vishwanath, Phys. Rev. Lett. {\bf 108}, 046602 (2012).
\bibitem{nagaosa} H. Isobe, and N. Nagaosa, Phys. Rev. B, {\bf 86},  165127 (2012).  
\bibitem{rosenstein} B. Rosenstein and M. Lewkowicz, Phys. Rev. B, {\bf 88}, 045108 (2013).
\bibitem{DSHW-FV} S. Das Sarma, and E. H. Hwang, Phys. Rev. B {\bf 87}, 045425 (2013).
\bibitem{felix} E. Miriani, and F. von Oppen, Phys. Rev. Lett. {\bf 100}, 076801 (2008); \emph{ibid} {\bf 100}, 249901(E) (2008).
\bibitem{peskin-schroeder} M. E. Peskin, and D. V. Schroeder, {\em An Introduction To Quantum Field Theory} (Addison-Wesley, Reading, MA, 1995).
\bibitem{roy-herbut-pseudo} B. Roy, and I. F. Herbut, Phys. Rev. B {\bf 88}, 045425 (2013).
\end{thebibliography}
\end{document}